\documentclass[twoside,twocolumn,9pt]{article}
\usepackage{extsizes}
\usepackage[super,sort&compress,comma]{natbib} 
\usepackage[version=3]{mhchem}
\usepackage[left=1.5cm, right=1.5cm, top=1.785cm, bottom=2.0cm]{geometry}
\usepackage{balance}
\usepackage{times,mathptmx}
\usepackage{sectsty}
\usepackage{graphicx} 
\usepackage{lastpage}
\usepackage[format=plain,justification=justified,singlelinecheck=false,font={stretch=1.125,small,sf},labelfont=bf,labelsep=space]{caption}
\usepackage{float}
\usepackage{fancyhdr}
\usepackage{fnpos}
\usepackage[english]{babel}
\addto{\captionsenglish}{%
  
}
\usepackage{array}
\usepackage{droidsans}
\usepackage{charter}
\usepackage[T1]{fontenc}
\usepackage[usenames,dvipsnames]{xcolor}
\usepackage{setspace}
\usepackage[compact]{titlesec}
\usepackage{hyperref}

\usepackage{epstopdf}

\definecolor{cream}{RGB}{222,217,201}

\begin{document}

\pagestyle{fancy}
\thispagestyle{plain}
\fancypagestyle{plain}{

\renewcommand{\headrulewidth}{0pt}
}

\makeFNbottom
\makeatletter
\renewcommand\LARGE{\@setfontsize\LARGE{15pt}{17}}
\renewcommand\Large{\@setfontsize\Large{12pt}{14}}
\renewcommand\large{\@setfontsize\large{10pt}{12}}
\renewcommand\footnotesize{\@setfontsize\footnotesize{7pt}{10}}
\makeatother

\renewcommand{\thefootnote}{\fnsymbol{footnote}}
\renewcommand\footnoterule{\vspace*{1pt}%
\color{cream}\hrule width 3.5in height 0.4pt \color{black}\vspace*{5pt}} 
\setcounter{secnumdepth}{5}

\makeatletter 
\renewcommand\@biblabel[1]{#1}            
\renewcommand\@makefntext[1]%
{\noindent\makebox[0pt][r]{\@thefnmark\,}#1}
\makeatother 
\renewcommand{\figurename}{\small{Fig.}~}
\sectionfont{\sffamily\Large}
\subsectionfont{\normalsize}
\subsubsectionfont{\bf}
\setstretch{1.125} 
\setlength{\skip\footins}{0.8cm}
\setlength{\footnotesep}{0.25cm}
\setlength{\jot}{10pt}
\titlespacing*{\section}{0pt}{4pt}{4pt}
\titlespacing*{\subsection}{0pt}{15pt}{1pt}

\renewcommand{\headrulewidth}{0pt} 
\renewcommand{\footrulewidth}{0pt}
\setlength{\arrayrulewidth}{1pt}
\setlength{\columnsep}{6.5mm}
\setlength\bibsep{1pt}

\makeatletter 
\newlength{\figrulesep} 
\setlength{\figrulesep}{0.5\textfloatsep} 

\newcommand{\topfigrule}{\vspace*{-1pt}%
\noindent{\color{cream}\rule[-\figrulesep]{\columnwidth}{1.5pt}} }

\newcommand{\botfigrule}{\vspace*{-2pt}%
\noindent{\color{cream}\rule[\figrulesep]{\columnwidth}{1.5pt}} }

\newcommand{\dblfigrule}{\vspace*{-1pt}%
\noindent{\color{cream}\rule[-\figrulesep]{\textwidth}{1.5pt}} }

\makeatother

\twocolumn[
  \begin{@twocolumnfalse}
\vspace{3cm}
\sffamily
\begin{tabular}{m{4.5cm} p{13.5cm} }


&\noindent\LARGE{\textbf{Evaporating microfluidic droplets: A tool to study pathogen viability in bioaerosols  $^\dag$}} \\
\vspace{0.3cm} & \vspace{0.3cm} \\

& \noindent\large Maheshwar Gopu$^{\ddagger}$, {Akanksha Agrawal$^{\ddagger}$, Raju Mukherjee$^*$, and Dileep Mampallil$^*$.} \\

\vspace{0.3cm} & \vspace{0.3cm} \\

&\noindent\normalsize{Pathogens in droplets on fomites and aerosols go through extreme physiochemical conditions, such as confinement and osmotic stress, due to evaporation.  Still, these droplets are the predominant transmission routes of many contagious diseases.  The biggest challenge for studying the survival mechanisms of pathogens in extreme conditions is closely observing them, especially in aerosols, due to the small droplet sizes, movements, and fast evaporative dynamics.  To mimic evaporating aerosols and microdroplets, we employ microfluidic emulsion droplets.  The presence of oil forms a microdroplet cluster with spatially gradient evaporation rates, which helps studying the impact of evaporation and its rate. With \textit{Escherichia coli}, we show that the viability is adversely affected by the evaporation and its rate. Our method can mimic bioaerosol evaporation with controlled number of cells inside. In general,  the droplets can act as microreactors with varying kinetics induced by different evaporation rates.  } 

\\

\end{tabular}

 \end{@twocolumnfalse} \vspace{0.6cm}

  ]

\renewcommand*\rmdefault{bch}\normalfont\upshape
\rmfamily
\section*{}
\vspace{-1cm}


\footnotetext{\textit{~Indian Institute of Science Education and Research Tirupati, Mangalam P. O. PIN 517507, Tirupati, AP, INDIA \newline E-mail: dileep.mampallil@iisertirupati.ac.in \newline E-mail: raju.mukherjee@iisertirupati.ac.in}}

\footnotetext{\ddag~These authors contributed equally to this work.}


\section{Introduction}

Bioaerosols and microdroplets on fomites transmit various diseases \cite{Jones2015, Tellier2019, Dbouk2020} even leading to epidemic outbreaks, for example, COVID-19. Transmission of many contagious diseases strongly depends upon the viability of pathogens within these droplets.

Aerosols and droplets on fomites undergo evaporation and sometimes condensation, changing the internal physiochemical environment, for example, osmolality and confinement. These changes affect the metabolism, physiology, and biological activity of the pathogens \cite{Tang2009, Lever2000, Lighthart1997}.  The self-secreted secondary metabolites of the microorganisms that stay within the droplets can also influence their viability. Nevertheless, successful transmission of many pathogens occurs without doubt.  In some cases, bacterial viability is not affected even after complete drying and rehydration of droplets \cite{Majee}. Thus, the transmission of viable and active pathogens  through microdroplets is a complex process. Understanding the complex process of survival in extreme conditions is important for disease management. 

There are several efforts to study the survival of microorganisms in aerosol droplets \cite{Haddrell2017}. A common method is bulk imaging of spray-generated microdroplets levitated against gravity by the usage of centrifugal forces in rotating drums \cite{Goldberg1958, Asgharian1992, Lever2000, Verreault2014}. Due to the increased gravitational force, it is not possible to levitate droplets of several tens of micrometers in size, which are the typical sizes of droplets generated by human activity, such as sneezing \cite{Han2013, ZhangPE2015}. To improve this scenario, levitating the aerosol droplets using an electrodynamic trap was also demonstrated \cite{Fernandez2019}. The drawback of these methods is that the levitated aerosol droplets are not suitable for close and prolonged observation for detailed studies. Due to the small sizes and fast dynamics,  imaging individual aerosol droplets is not practical with existing techniques.   Moreover, laboratory generation of aerosols, often by nebulization, also compromises the survival of pathogens in the droplets \cite{Zhen2014}. Nebulization cannot also generate droplets with controllable sizes. Another hurdle is the fast evaporation of these tiny droplets. It prevents observing them for a considerable duration of the study. In short, there is a need for a softer generation of aerosol-sized droplets, control of their evaporation, and close observation of microorganisms inside the individual aerosol droplets.

Evaporation of individual and clusters of droplets \cite{Hu2002, Brutin2018, Laghezza2016, Hatte2019, Wen2019} was previously proposed as a method for studying individual aerosols \cite{Netz2020, Dombrovsky, Grinberg, Seyfert2021}. However, these large droplets, placed directly in the ambient air, offered no confinement effects. Therefore, large droplets do not represent conditions inside the microdroplets on fomites or aerosols.  Introducing microbes and controlling their number inside these large droplets are also difficult. 

\begin{center}
	\begin{figure*}[ht]
		\centering
		\includegraphics[scale=1.1]{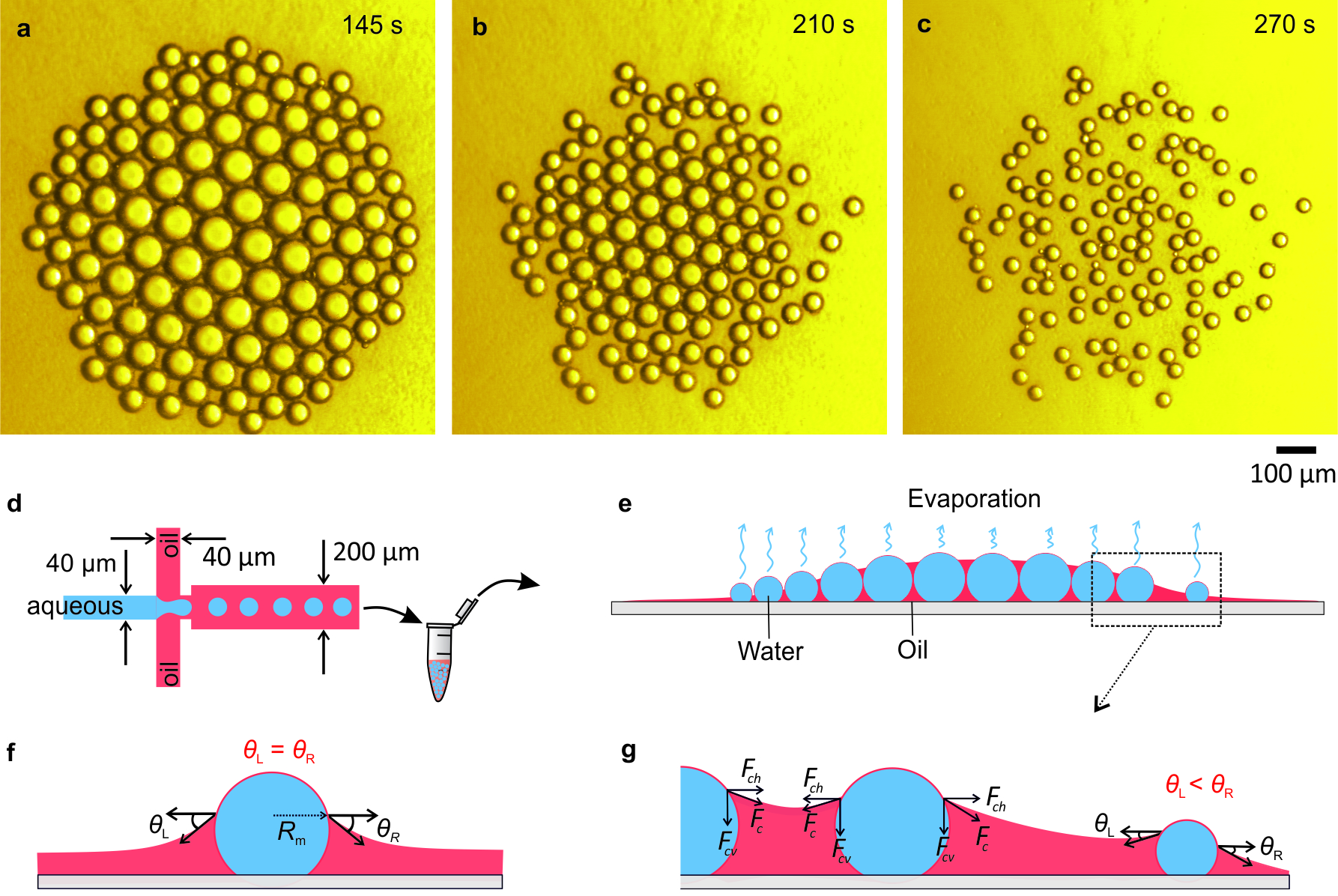}
		\caption{\textbf{Principle: gradient evaporation} \textbf{a-c} Images of the cluster of microdroplets captured at different time points during the evaporation (before 145 s, the size of the cluster was larger than the viewed area). The size of aqueous microdroplets decreases faster near the cluster edge.  \textbf{d} Illustrations of the drop generation in a flow-focusing device. \textbf{e}  The oil drop, partially overlapping the microdroplets, creates a spatially varying evaporation rate (represented by the wavy arrows). \textbf{f, g} Illustrations show the capillary forces due to the oil meniscus around a single droplet (\textbf{f}) and multiple droplets (\textbf{g}). Asymmetry in the meniscus shape produces a net horizontal component of the capillary force towards the cluster center. }\label{FigMain}
	\end{figure*}
\end{center}

Here, we propose microfluidics as a platform to overcome the difficulties in observing inside the individual aerosol microdroplets, with advantages of studying even single cells. The microfluidic method of generating nanoliter volume droplets with high uniformity is a well-established process \cite{Zhu2017}. Biocompatibility  of these microdroplets in the ambient oil medium was well demonstrated in various cases, for example, single-cell analysis \cite{Shinde2018, Gao2019, Matua2020}, drug discovery \cite{Neuzil2012, Shembekar2016}, and directed evolution \cite{Agresti2010}, to list a few. When an emulsion drop is deposited on a substrate, the aqueous microdroplets self-assemble to form an array (Fig.~\ref{FigMain}a). Our method exploits the slow and spatially gradient evaporation of the array of microdroplets.  Particularly,  the spatially varying evaporation rate helps us to compare the effects of different rates of increase of chemical contents, especially salt. As a proof of concept, we exposed \textit{Escherichia coli} to different rates of increase in salt in the microdroplet cluster and show that higher rates compromised their viability. Being droplet microfluidics a  well established and bio-friendly platform, our report can quickly urge other scientists to use such a system for exploring biology inside aerosols.

\section{Experiments}  

\subsection*{Chemicals, Bacterial Strains and Culture} 
Propidium Iodide was purchased from Invitrogen. All other chemicals were purchased from Sigma Aldrich. MilliQ water was used where necessary.

For fluorescence experiments, Tris(2,2'-bipyridine)ruthenium(II) hexafluorophosphate (hereafter called ruthenium bipyridine) was mixed in deionised water at an initial concentration of 4 mM. The molecule has its excitation and emission at blue and red wavelengths, respectively.

Artificial saliva was prepared by mixing mucin from the porcine stomach (Sigma Aldrich) at 1\,g/L, and NaCl at 3\,g/L in deionised water.

We chose a high copy number plasmid (100-300) pSB1C3 with SYFP2 gene (a brighter and photostable variant of green fluorescent protein)\cite{Kremers2006} expressed from a T7 promoter. This promoter has a leaky expression in BL-21 (DE3) strain of \textit{Escherichia coli} leading to a cytoplasmic production of green fluorescence without any induction.

Lab-prepared chemically competent \textit{E. coli} BL-21 DE3 strain was transformed with the above-mentioned pSB1C3 plasmid with chloramphenicol as the antibiotic marker. The transformed cells were then plated on Luria-Bertani (LB) agar supplemented with chloramphenicol (35 $\mu$g ml-1) and cultured overnight at 37 $^\circ$C. A single colony was inoculated in LB media supplemented with chloramphenicol (35 $\mu$g/ml) and incubated at 37 $^\circ$C with shaking at 200 rpm till the culture reached an OD600 of 0.1. The cells were then washed with phosphate-buffered saline (PBS) and resuspended in PBS at 100x dilution. Propidium Iodide was added at a final concentration of 1.5 $\mu$M to the bacterial suspension and used for generating bacteria encapsulated droplets. Propidium Iodide was used for differentiating live cells with intact cell envelope (green) from the ones with compromised cell wall (green + red) \cite{Boulos1999}.

\subsection*{Device Fabrication $\&$ Droplet Generation}
 Droplets were produced using a flow-focusing microfluidic device. The device was fabricated with PDMS using the soft-lithography technique. The channel design and the dimensions are shown in Fig 1(e). To stabilize the emulsion, Span 80 was added to the oil, hexadecane, at a concentration of (a) 2.8\% (v/v) for water  (b) 2.7\% (v/v) for PBS when encapsulated the bacteria.

We used syringe pumps (Harvard apparatus pump-11 elite) to pump the fluids into the fabricated PDMS device. The flow rate for dispersed (aqueous) and continuous medium (oil) were kept at 0.2 $\mu$l/min and 0.2 $\mu$l/min, respectively. The generated emulsion was collected in a micro-centrifuge tube. For further experiments, a small volume of emulsion ranging from 0.5 to 2 $\mu$l was placed on a glass slide that was cleaned using acetone, ethanol, and water followed by spin-drying.

To view the oil layer around the microdroplets, we used the fluorescent dye Nile Red at the concentration 0.3 $\mu$M added to Hexadecane + 2.8\% (v/v) Span 80.
All the emulsion evaporation experiments were conducted at room temperature: 25 $^\circ$C with relative humidity (45\% to 50\%).

\subsection*{Imaging and Analysis}
 The emulsion drop placed on the glass slide was imaged using (a) an inverted fluorescence microscope (Nikon Eclipse Ti-U) and the camera (Nikon DS-U3) at regular intervals (typically 1 s), (b) Confocal system (Leica SP8). The $\lambda_{ex}$ : $\lambda_{em}$ spectrum were as follows (a) Nile red, 530 nm : 635 nm; (b) PI, 535 nm : 617 nm; (c) SYFP2, 515 nm : 527 nm \cite{Kremers2006}. The captured images were processed and analyzed using the software: ImageJ (NIH) and CellProfiler.

\section{Results }  

\subsection{Principle}

Our strategy is to place a microfluidic emulsion drop outside the microfluidic system in ambient air and allow the aqueous microdroplets in the emulsion to evaporate.  Immediately after placing the emulsion drop, the aqueous microdroplets settle onto the glass surface as the oil is lighter than water.  This configuration is like a large oil drop containing a cluster of aqueous microdroplets. We illustrate this scenario in Fig.~\ref{FigMain}e. The main principle of our method is that the microdroplets  are partially exposed to evaporate depending upon their position in the oil drop. The radial decrease of the oil height results in an increased exposing of microdroplets sitting towards the cluster edge. 

On top of changing the degree of evaporation, the presence of the oil layer in between the microdroplets (see Fig.~S1 in the Supplementary Information) has also a role in keeping the droplets clustered. To understand this aspect, consider a single aqueous droplet partially immersed in an oil layer. The radially symmetric oil meniscus formed around the droplet produces no net horizontal component of capillary force Fig.~\ref{FigMain}f. However, the vertical component pulls the droplet towards the substrate.  This scenario changes when the menisci at both sides of a microdroplet have different slopes, for example, due to another droplet nearby, as illustrated in Fig.~\ref{FigMain}g. Here, a net horizontal force is generated, pulling the microdroplets closer to each other. This net horizontal component of the force acting on a single droplet is given by 

\begin{equation}
F_{ch} \sim  R_m \,\gamma_{oa} (\cos\theta_L - \cos\theta_R), \label{Eq1}
\end{equation}

where $\gamma_{oa}$ is the surface tension of the oil, $R_m$ is the radius of the meniscus, and $\theta_L$ and $\theta_R$ are the angles subtended by the meniscus at both sides of the microdroplet as shown in Fig.~1f. In the case of a cluster of microdroplets, the radially inward capillary attraction balanced by the outward push due to the capillary deformation of droplets brings an equilibrium. Such capillary interactions are reported, for example, in the cases of mimicking the cell-cell adhesion pressure using functionalized emulsion droplets \cite{Pontani2012}, capillary orbiting of silicone oil droplets dispersed on liquid nitrogen surface \cite{Gauthier2019}, aqueous droplets placed closely on oil-coated surfaces \cite{Sun2019, Boreyko2014, Kajiya2016}, droplets sliding on gels \cite{Karpitschka2016}, and colloidal particles suspended at water-air interface \cite{Vassileva2005}. 

The microdroplet cluster, partially immersed in the oil, evaporates with specific features. (i) First, the size of the microdroplets decreases in a non-uniform fashion. The outer microdroplets shrink faster than the ones at the middle of the cluster, as seen in Fig.~\ref{FigMain}a. It means that within a cluster, we have droplets evaporating at different rates, producing different rates of increase of chemical/salt stress. (ii) Second, the overall evaporation rate of the microdroplets is low, taking several tens of minutes for the complete evaporation. It gives an ample amount of time for monitoring inside the droplet. Moreover, by altering the radially outward flow of the oil or changing the ambient saturation level, we can tune the overall evaporation time.

\begin{center}
	\begin{figure*}[ht]  
		\centering
		\includegraphics[scale=1.0]{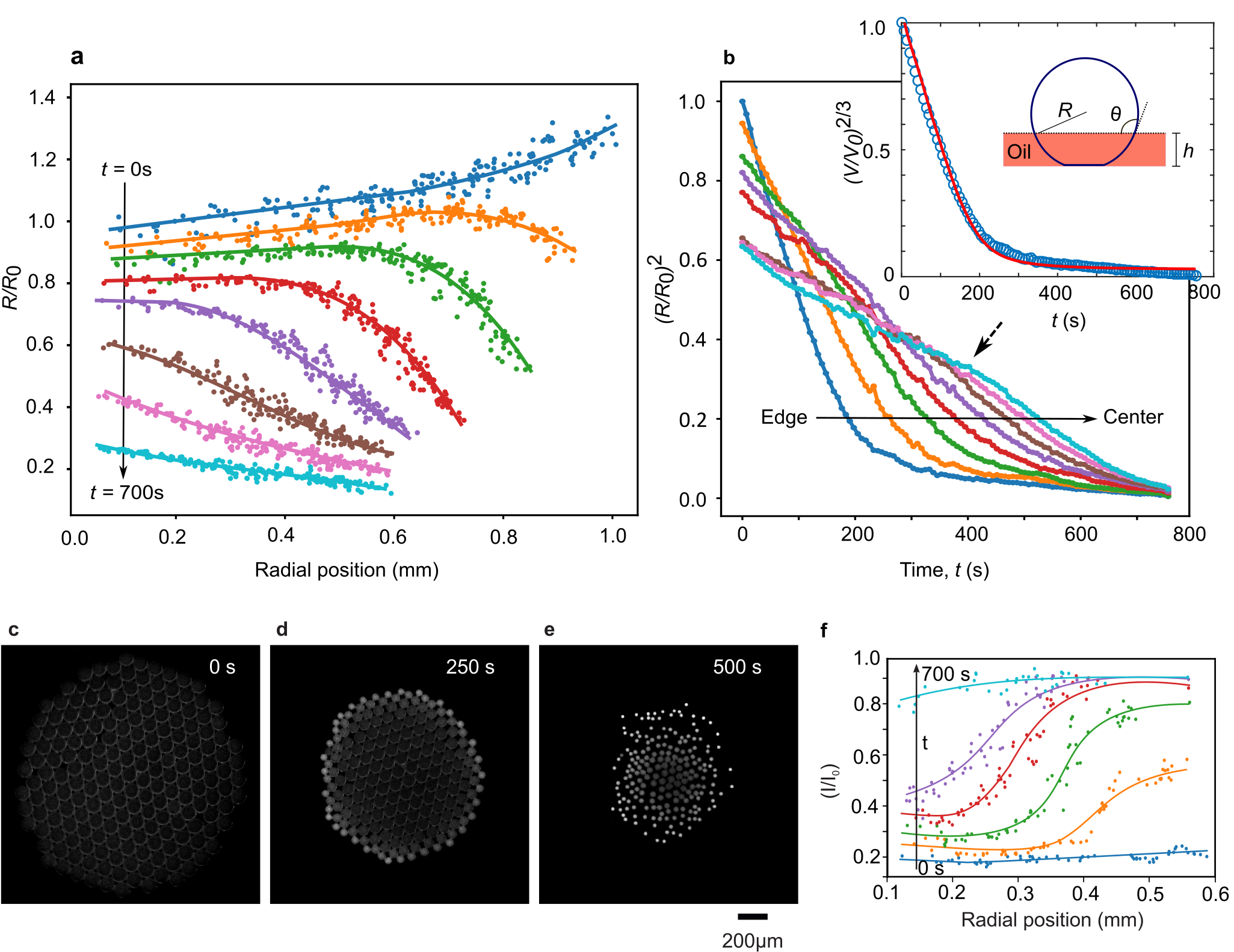}
		\caption{\textbf{Size gradient of droplets.} \textbf{a} The radius distribution of the microdroplets from the center to the edge of the cluster. The curves corresponds to $t$ = 0, 100, 200 ... 700 s.  Each data point represents a single microdroplet (see SI). The radius is normalized by the initial value (50 $\mu$m). The solid lines are guide to the eyes.  \textbf{b}  Normalised area versus time for  individual droplets at different radial positions in the cluster.  The curves decrease linearly (Eq.~\ref{eq4}) and flatten off as the droplet diameter approaches the thickness of the oil layer. The inset shows the theoretical model curve fitted to the data of the edge droplet. The oil height $h$ and angle $\theta$ are shown in the illustration.  The initial emulsion drop volume was 0.5 $\mu$l. \textbf{c-e} Images of evaporating droplet-cluster with fluorescing ruthenium bipyridine in the liquid. The edge droplets start to fluoresce at the beginning followed by the inner droplets as quantified in (\textbf{f)}. The solid lines in (\textbf{f}) are guides to the eyes.  }   \label{FigGrad}
	\end{figure*}
\end{center}

\subsection{Gradient in the size of the microdroplets}

Immediately after placing the emulsion drop on the substrate, the microdroplets at the cluster edge seem larger than those at the middle of the cluster (not seen in Fig.~1 but in Fig.~S2 in the Supplementary Information).  This aspect is also quantified in the curve corresponding to $t$ = 0 in Fig.~\ref{FigGrad}a, where the size of the droplet is plotted as a function of the radial distance from the center of the cluster. The droplets towards the edge look bigger because they are less compressed by the capillary attraction, unlike the ones inside the cluster. Although the droplets in the cluster have the same volume, viewing them from the bottom, the inner ones appear to have a smaller diameter.  

The size of the microdroplets at the edge of the cluster starts to decrease at a faster rate compared to those inside the cluster (see the images Fig.~\ref{FigMain}a-d and Fig.~\ref{FigGrad} (a)). For example, as seen in Fig.~\ref{FigGrad}(a), at $t = $ 100 s, the droplets near the edge became smaller than the ones towards the cluster center. With time, the evaporation rate increases for more droplets within the cluster.

In Fig.~\ref{FigGrad}(b), we plotted the square of the normalized radius versus time for individual microdroplets at different radial positions in the cluster. Here also we observe that the droplets near the edge start to decrease their size at a fast rate early. In the fast evaporation region, the curves show a linear behavior. Nevertheless, with time, the linear behaviour changes and the curves flatten off. It implies that the evaporation stops gradually.  We expect this behavior as the droplets stop evaporating once their diameter becomes smaller than the thickness of the oil layer there.

The middle droplets evaporate slowly followed by a relatively increased rate. The point at which the rate increases is marked by the arrows in Fig.~\ref{FigGrad}(b). At this point, the oil layer around the middle droplets flows out exposing them for faster evaporation. The oil could flow freely as the outer droplets became smaller creating space. Overall we can see that, in Fig.~\ref{FigGrad}(b), the slope of the linear region decreases for droplets sitting towards middle of the cluster.  This aspect supports a thicker oil layer at the middle, as explained in the case of Fig.~\ref{FigGrad}a. The thicker oil layer at the middle decreases the exposed area of droplets for evaporation.  

Finally, we show the concentration increase of chemical components inside the droplets at different rates across the cluster.  We use ruthenium bipyridine in the water microdroplets and measure the fluorescence intensity. The intensity is increasing from the edges at different rates as shown in Fig.\ref{FigGrad}(c-f). With time, the intensity of inner droplets increases as the concentration of ruthenium bipyridine increases in them.

\subsection{Modelling the evaporation}

We can understand the curves in Fig.~\ref{FigGrad}(b) using a model where the evaporation is taken as a diffusion process driven by the difference in the vapor concentration near the droplet surface and far away \cite{Hu2002, Erbil2012}.  The change in the mass of the droplet is given as
 
\begin{equation}\label{eq2}
\frac{dm}{dt} = - D \int\nabla c.dS,
\end{equation} 

where $D$ is the diffusion coefficient, $c$ is the vapor concentration, and $dS$ is the area element. Eq.\ref{eq2} can be written in terms of the volume, $V$, as \cite{Erbil2012},

\begin{equation}\label{eq3}
\frac{dV}{dt} = - \frac{4\pi R\,D (c_s-c_{\infty})}{\rho}  f(\theta),
\end{equation} 

where, $c_s$ and $c_{\infty}$ are the vapour concentrations at drop surface and infinity, respectively,  $\rho$ is the density of the fluid, and $R$ is the radius of a microdroplet. The function $f(\theta) = -cos\theta/2 log_e(1-cos\theta)$  accounts for the vapour distribution across the curved drop \cite{Erbil2012}. We can rewrite the radius of the drop as $R = 3 V/\beta\pi$, with $\beta  = (1-\cos\theta_s)^2(2+\cos\theta_s)$, where $\theta_s$ is the contact angle of the  droplet on the glass surface (with ambient oil).  Integrating Eq.\ref{eq3}, we obtain,

\begin{equation}\label{eq4}
V^{2/3}(t) = V_{0}^{2/3} -  \frac{8 \pi D  (c_s - c_{\infty}) }{3 \rho} \left( \frac{3}{\pi\beta} \right)^{1/3} k\,f(\theta)\,t,  
\end{equation}

where $k$ is an empirical parameter that we additionally included to account for the overall slower evaporation compared to a bare droplet in ambient air. We assume a droplet partially immersed in an oil layer of uniform and constant thickness $h$ (see the schematic in the inset). We take $\theta_s$ as constant at 150$^\circ$ (measured for a 1 $\mu$l drop) resulting $\beta$ = 3.95. The evaporation continues until the drop diameter becomes equal to the height of the oil. 

For a partially immersed water droplet in oil, the Neumann angles at water-gas, oil-gas, and water-oil interface change with the relative heights of the droplet and oil \cite{Semprebon}. Therefore, the value of $\theta$ decreases with time as the droplet height approaches that of the oil \cite{Semprebon}.  For simplicity, we took $\theta_L = \theta_R = \theta$. In Eq.\ref{eq4}, purely geometrically, we approximate the time varying contact angle as $\theta(t) = \pi - \cos^{-1}(1-h/R(t))$, where $R(t)$ is obtained from the total droplet volume (see SI). 

To obtain the model curve, we took the value of $c_s$  = 0.023  Kg/m$^3$,  $c_{\infty}$ = 0.0092 Kg/m$^3$, and $D$ = 24 $\times$ 10$^{-6}$ m$^2$/s, at 25 $^\circ$C \cite{Hu2002}. The initial droplet volume $V_0$ was 0.5 nL (initial radius $R_0$ = 50 $\mu$m).  To test the model, we fitted Eq.\ref{eq4} to the experimental curve corresponding to the edge droplet in Fig.\ref{FigGrad}(b). The model is plotted in the inset of Fig.\ref{FigGrad}(b). We adjusted the values of $h$ = 17 $\mu$m and the final droplet volume $V_f$ = $V_0/$230. These values determine the shape of the linear to flat transition region and position of the tail of the curve. 

The slope of the initial linear region was optimised by adjusting the value of $k$ to 0.044. Thus, empirical factor $k$ accounts for the slow evaporation of the droplets. In the ambient conditions in our laboratory, an isolated droplet of typical volume 0.5 nL takes only less than 10 seconds for complete evaporation. The clustering of droplets does not change the evaporation time of individual droplets dramatically \cite{Carrier2016}. However, in our experiments,  an individual microdroplet of typical volume 0.5 nL takes  around 700 to 900 seconds as inferred from Fig.~\ref{FigGrad}. As the droplets in the cluster evaporates at different rate, the central droplets have longer evaporation time, which also depends on the cluster size (Fig.Sx).

There are two reasons for the slow evaporation of microdroplets \cite{Gao2019, LiPNAS2020}:  (i) First, the area ($\sim \pi R_m^2$) of the microdroplet exposed to the air is small as the droplets are partially immersed in the oil layer.  (ii)  Second, the presence of a thin oil film on the exposed surface of the microdroplets.  A thin layer of oil cloaks the water droplets for a positive value of the spreading coefficient, $S_{ow}=\gamma_{wa}-\gamma_{oa}-\gamma_{ow}$, where $\gamma$ is the interfacial tension between the two phases denoted by subscripts $w$ (water), $o$ (oil), and $a$ (air) \cite{Smith2013, Schellenberger2015, Daniel2017}. With $\gamma_{wa}=$ 73 mN m$^{-1}$,  $\gamma_{oa}=$ 28 mN m$^{-1}$ , and $\gamma_{ow} = $  2.5 mN m$^{-1}$ \cite{Politova2017}, we obtain $S_{ow} > $ 0. When hydrostatic effects are neglected, the thickness of the cloaked oil film can be obtained from the balance between the capillary pressure due to curvature and the disjoining pressure between water-oil and oil-air interfaces \cite{Schellenberger2015}. The thickness of oil film $b$ is given by 

\begin{equation}
b = \left(\frac{A\, R_0}{12 \pi \gamma_{oa}}\right)^{1/3},
\end{equation}

where $A$ is the Hamaker constant, which can be estimated by using the non-retarded Hamaker constant in Lifshitz theory as in ref.\cite{Daniel2017}. We estimated the value of the Hamaker constant as 6 $\times 10^{-21}$~J (Supplementary Information) and obtained the equilibrium thickness of the oil film as $\approx$ 10 nm.
The water molecules can still diffuse through this cloaked oil film and evaporate \cite{Gao2019, LiPNAS2020}.  

Our model did not account for any dynamic changes in the oil height and the cloaked oil film thickness. Both the factors must influence the rate of decrease of the droplet size. Including such aspects requires further understanding on the oil layer thickness between the droplets and dynamics of the clocked oil film, for example, using interferometric methods \cite{Marklund2001, Daniel2017, Kreder2018}. 

\begin{center}
	\begin{figure}[h] 
		\centering
		\includegraphics[scale=1]{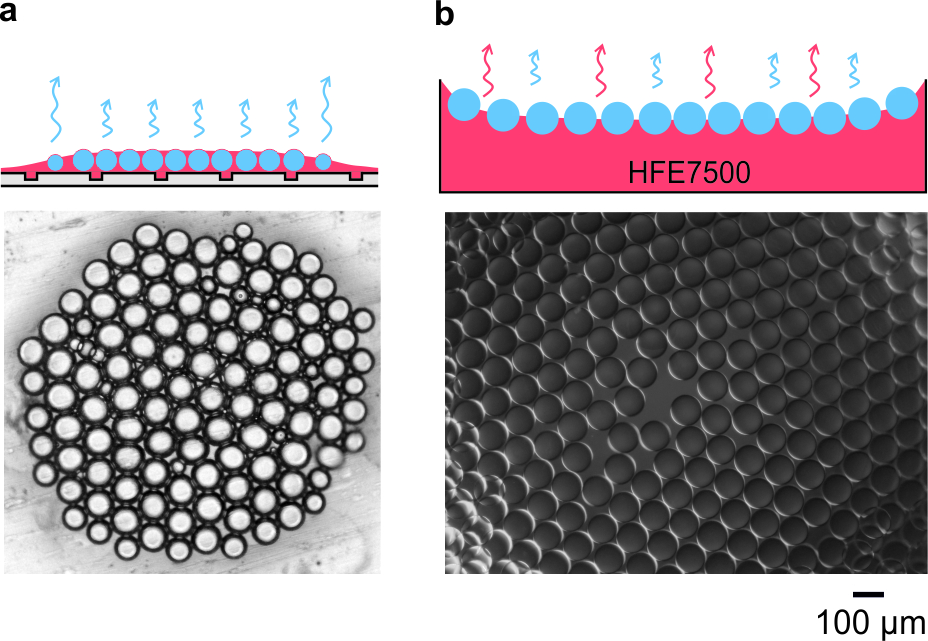}
		\caption{\textbf{Importance of oil.} \textbf{a} A lighter oil is required to partially expose the denser aqueous droplets. But if the oil flows out from the cluster, for example, by micro-scratches on the glass, no  gradient in microdroplet size is formed.  \textbf{b} A denser oil (HFE 7500) keeps all the microdroplets on the surface without forming any gradient evaporation.  }   \label{FigOIL}
	\end{figure}
\end{center}

\subsection{Importance of the oil and its properties}

Spatial gradients in the evaporation rate are reported in arrays of droplets \cite{Laghezza2016, Hatte2019, Wen2019}. For example, in the case of two evaporating droplets placed nearby,  the dense vapour between them decreases the evaporation rate at the mutually facing sides. On the opposite side, the evaporation rate is relatively higher. Similarly, in an array of droplets, the ones at the edge of the array evaporate faster. 

In our experiments, the influence of the different vapour densities on the evaporation rates will be less important as there is an oil layer present between the droplets. The mechanism for the gradient evaporation of the microdroplets  is their partial exposure as illustrated in Fig.~\ref{FigMain}e.  As the microdroplet cluster immersed in the oil is partially exposed as the oil drop is thinner towards the edge.  

To confirm the influence of the oil layer in creating the size gradient in the cluster, we introduced rough channels on the glass slide using sandpaper. The channels allowed the oil to flow out from the middle regions of the cluster. Interestingly, in this case, the evaporation did not create any noticeable size gradient of the microdroplets (Fig.\ref{FigOIL}(a)). When the oil flows out from the middle of the cluster, the radial height variation of the oil drop disappears. Therefore, all the microdroplets in the cluster evaporate at a similar rate, producing no size gradience.

We tested a scenario of droplets floating on the oil surface using the oil HFE 7500 (with surfactant), having a density of 1.6 g/cm$^3$. In this case, as shown in Fig.\ref{FigOIL}(b), the microdroplets were floating on the surface of the oil. The oil was taken in a container and not placed on a bare glass slide due to the very low surface tension of the oil (5 mN/m). If placed on the glass slide, the oil does not stay as a drop but completely spreads. The microdroplets on the oil surface evaporate without producing any size gradient as every droplet is exposed to the ambient air at a similar degree. Another difficulty with this system is the evaporation of the oil itself. The boiling point of HFE 7500 is 128 $^\circ$C, which is close to that of water. Therefore, both the oil and the aqueous droplets evaporate. Thus, a denser and evaporating oil is not a choice to create size gradient evaporation of the droplets.  

\subsection{Evaporation of saliva-like microdroplets }

\begin{center}
	\begin{figure}[h] 
		\centering
		\includegraphics[scale=1.05]{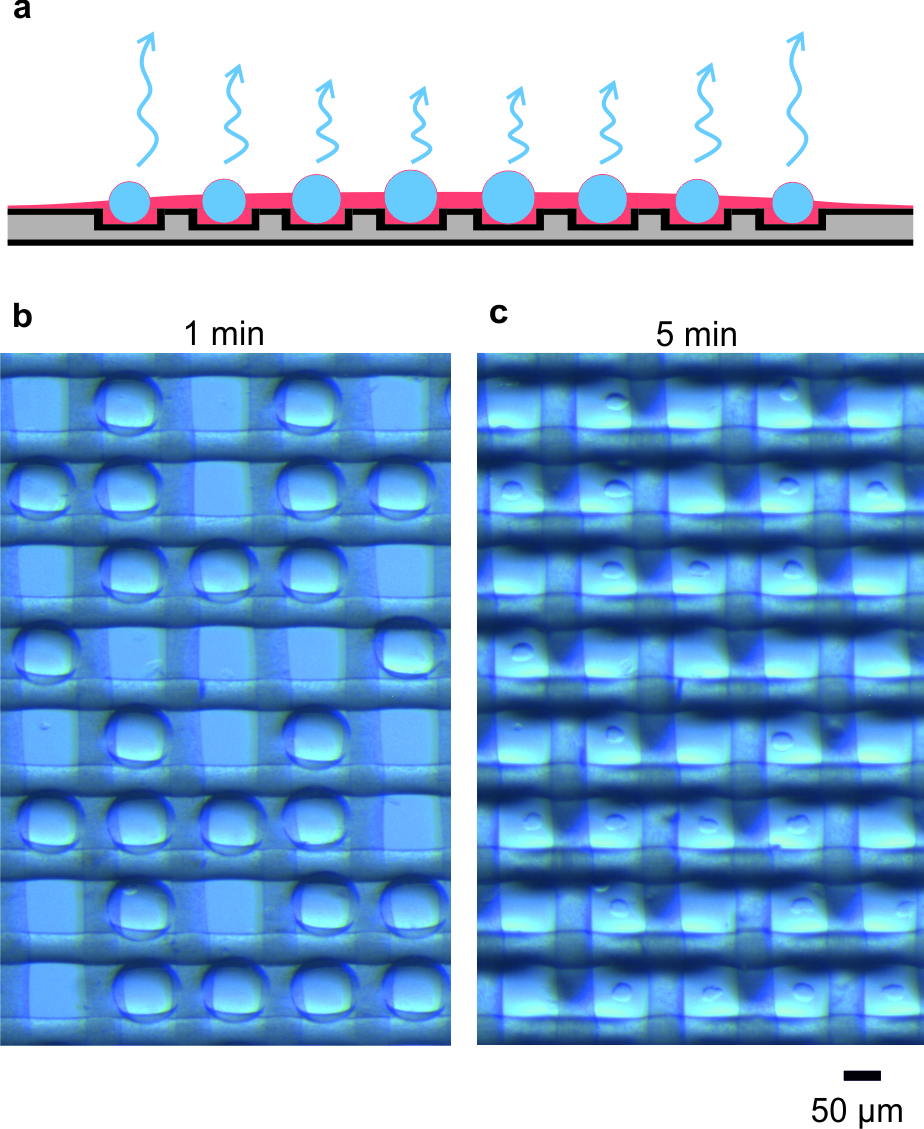}
		\caption{\textbf{Saliva-like samples.} \textbf{a} Illustration of the concept of physical separation of the microdroplets to prevent coalescence. \textbf{b, c} Artificial saliva droplets trapped on a nylon micromesh evaporate without coalescence.} \label{FigMESH}
	\end{figure}
\end{center}

Aerosols and microdroplets on fomites can be of saliva or other complex fluids. We  generated microdroplets using artificial saliva and real saliva samples (donated by one of the authors). Although droplet generation with artificial/real saliva is feasible, presence of components that decreases the surface tension is a hurdle in our method. We observed that the microdroplets started to coalesce to form bigger droplets after the onset of evaporation (Supplementary Fig. X). We overcame this difficulty by increasing the separation between the individual microdroplets using physical constraints. We used a small piece ($\sim$ 3 mm$\times$3 mm) of a nylon sieve fixed on a glass slide with tapes. When the emulsion drop was placed, the microdroplets fell in the rectangular meshes of each size $\sim$100 $\mu$m$\times$100 $\mu$m as shown in Fig.\ref{FigMESH}(f-g). Thus, physical separation allows studying droplets with lower surface tension, for example, saliva.

\begin{center}
	\begin{figure*}[h]
		\centering
		\includegraphics[scale=1]{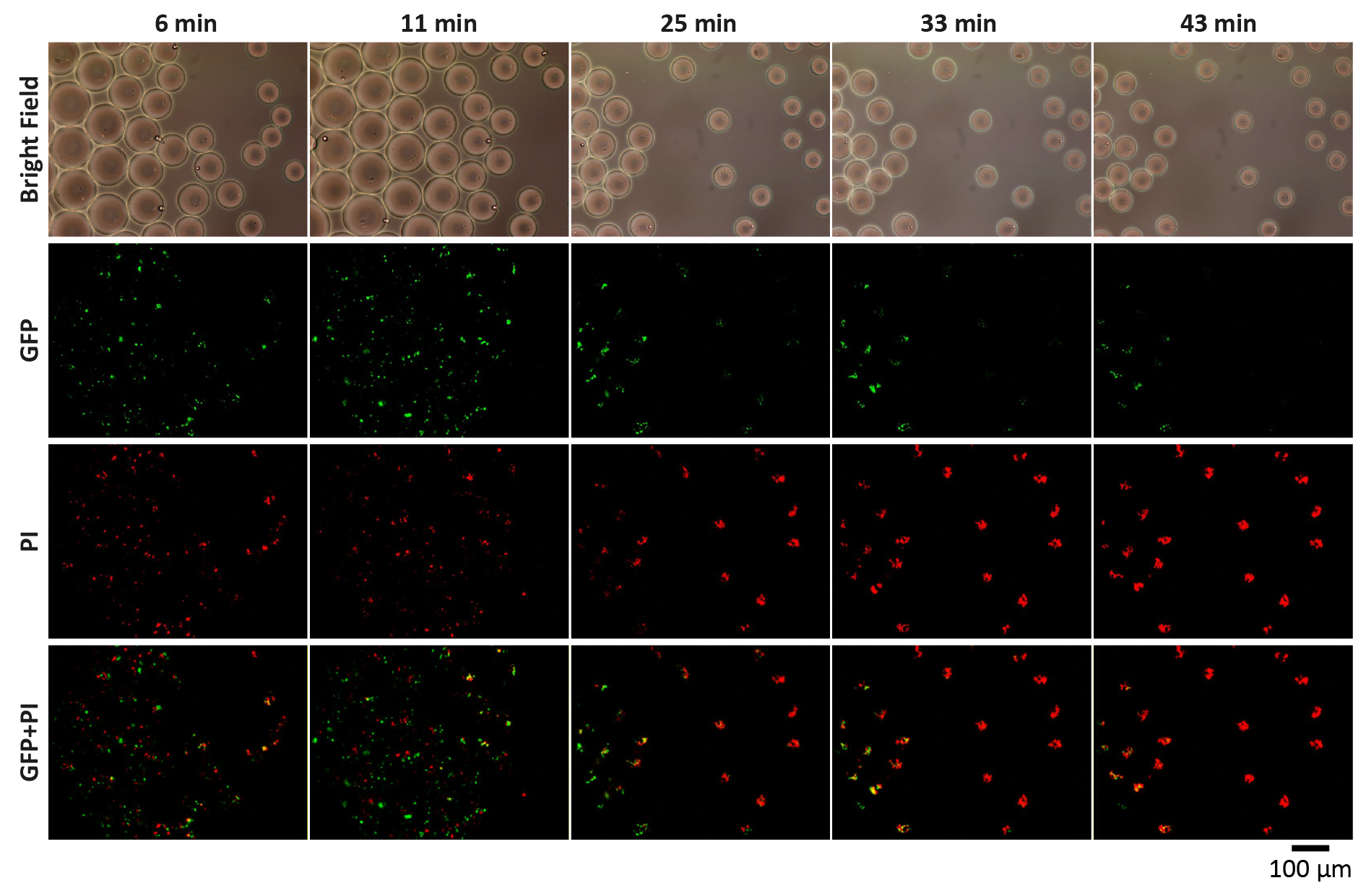}
		\caption{\textbf{\textit{ E. coli} encapsulated in the microdroplets.}  Time-lapse images of \textit{E.coli} constitutively expressing green fluorescent protein and stained with propidium iodide upon death.} \label{FigB1}
	\end{figure*}
\end{center}

\begin{center}
	\begin{figure*}[h]
		\centering
		\includegraphics[scale=1]{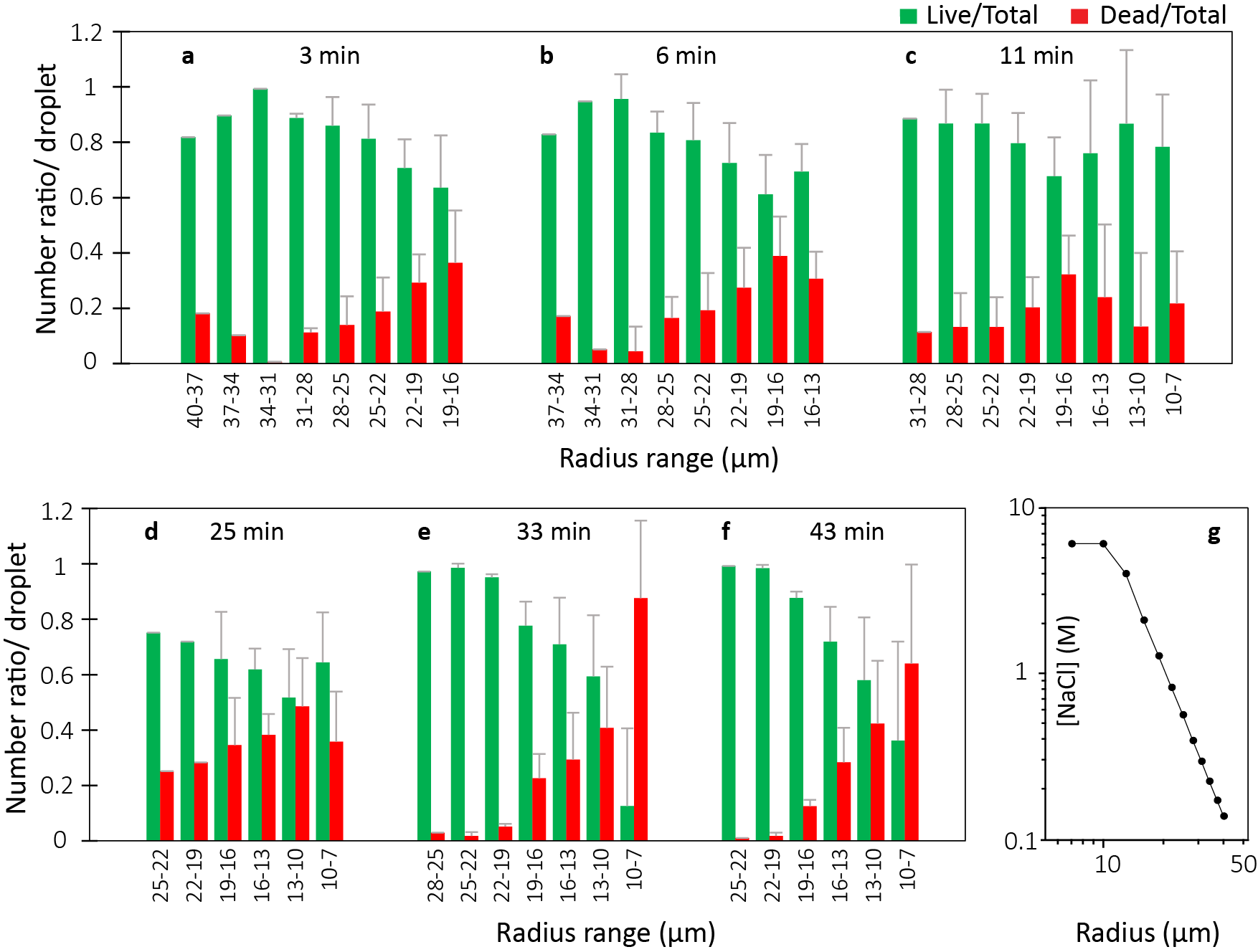}
		\caption{\textbf{Viability of \textit{E. coli} inside the microdroplets}.  \textbf{a - f} Fraction of live and dead (PI stained) \textit{E. coli} distributed according to the droplet size range at different time points. The error bars are the standard deviation. Slow evaporation leads to higher salt tolerance. \textbf{g} Change in NaCl concentration as a function of droplet size. The initial radius of the droplets was 40 $\mu$m.  The initial NaCl concentration was 173 mM.} \label{FigB2}
	\end{figure*}
\end{center}

\subsection{Bacteria inside the microdroplets}

It is interesting to understand how bacteria adapt to the changing conditions inside microdroplets during evaporation. We encapsulated bacteria in the droplets and studied their viability as the droplets evaporated at different rates.

As the droplets evaporate, the salt concentration gradually increases from 0.137 M to a maximum of 6.1 M (saturation concentration). We could see crystal formation within the small droplets as the concentration reaches the saturation concentration. Beyond the saturation concentration, it is reported that the evaporation rate decreases particularly at high humidity due to the hygroscopic effects of the salt \cite{Seyfert2021}. Similar effects may occur in our droplets as well, contributing to further the slowing down of the evaporation. As the salt concentration approaches the saturation value,  we expect that the increasing osmotic stress on the encapsulated bacteria affects their viability. To check the viability of the bacteria in the droplets, we added propidium iodide (PI) to the culture that was resuspended in PBS. PI stains the DNA in a bacterium of the damaged cell membrane. PI is impermeable across the intact membrane of a viable cell. PI staining is a good measure of the viability of bacterial cells except in the case of adherent cells in biofilms due to the presence of extracellular nucleic acids \cite{Rosenberg2019}. 
	
In our experiments, the PI concentration increased from 1.5 $\mu$M to about 280 $\mu$M (Table 1 in the Supplementary Information). To ensure that the increasing concentration of PI did not affect the cell viability, we performed a microtitre plate assay, where we incubated \textit{E. coli} cells for 2 hrs at 37 $^\circ$C in a static condition at various PI concentrations. Our fluorescence imaging showed that even at the highest concentration of PI (280 $\mu$M), only about 10\% of the cells got stained after 2 hrs of incubation (Fig.~S4 in the Supplementary Information). Thus, we concluded that even the highest PI concentrations did not interfere with the cell viability within our experimental duration. It implies that the membrane integrity was affected by the increasing osmolytes in the droplet and not by the increasing PI concentration. 

In Fig.~\ref{FigB1}, we show the images of a region of the microdroplet cluster. Initially, the droplets contained more live bacteria indicated by the green fluorescent signals. With time, we see that the droplet size decreases, and more red spots appear within the droplets. These red stains are due to PI reaching inside the cells indicating disintegration of the cell envelope. 

In Fig.~\ref{FigB2}, we quantify the number of dead and live cells as the droplets decrease their size. Overall, we have analyzed a total of 3480 bacteria encapsulated at several droplets. In Fig.~\ref{FigB2}, we plotted the \textit{dead/total} and \textit{live/total} ratio per droplet. The data is from different sets of droplets at different regions of the cluster, continuously observed as a function of time. We analyzed several such droplet clusters.
 At any time, we have droplets of different sizes in the cluster due to the gradient evaporation rate. 
 
Immediately at the beginning of the evaporation, the droplets were large, falling into the range of 40-37 $\mu$m as seen in Fig.~\ref{FigB2}a,  corresponding to time 3 minutes. The smallest range of droplet size that we imaged was 10 – 7 $\mu$m. Thus, in every panel in Fig.~\ref{FigB2}, droplets disappear when they are smaller than 7 $\mu$m. This size limit is due to our difficulty in imaging, and not a limit of the proposed technique itself.
 
It is important to note that, in Fig.~\ref{FigB2}, the same radius range of droplets at different panels took different times to reach those sizes. For example, the droplets with 19-16 $\mu$m in panel (a) took only 3 minutes to reach that range from the initial radius of 40 $\mu$m. On the other hand, some droplets took 43 minutes to reach the range 19-16 $\mu$m as seen in panel (f). Thus, the droplets represented in panel (f) evaporated slowly. In short, the same radius range of droplets at a different time (panels) experienced a different rate of evaporation.

In general, we can see that the smaller droplets have more dead bacteria in them. On top of that, we also see a correlation between the evaporation rate and the number of dead bacteria.  For example,  in Fig.~\ref{FigB2}, on panels (a) and (b), we see an increase in the number of dead bacteria in droplets of radius below 25 $\mu$m. These droplets, situated near the edge of the cluster, were initially having a radius around 40 $\mu$m. Their size decreased to below 25 $\mu$m taking just 3 and 6 minutes, respectively. It should be also noted that the edge droplets decrease their size sharply at the initial stage of evaporation (Fig.~\ref{FigGrad}b).  

After an initial increase in the number of dead cells, we do not see any substantial increase in them in smaller droplets. The maximum number of dead cells is roughly below 40\% of the total. Only after a long time, at 33 and 43 minutes, we can find a relatively larger number of dead cells in the smaller droplets. It means that after the initial increase in the dead bacteria, any further increase is slow.  For example, we consider droplets with a broad radius range of 25-16 $\mu$m. We can compare the number of dead cells in the case of fast evaporation and slow evaporation, i.e.,  time range of 6-11 min and 33-43 min, respectively.  We observe that the number of dead cells was surprisingly low for slow evaporation taking 33 - 43 min despite the salt concentration was being similar in the droplets. The number of dead cells was higher for faster evaporation taking only 6-11 min. This observation makes us assume that faster evaporation creates more adverse effects than a slower one.

\section{Discussion}

Many infectious diseases, for example, human tuberculosis, legionellosis, and many viral infections are transmitted by aerosols and microdroplets on surfaces \cite{Tellier2019}. Understanding the viability of microorganisms inside aerosols and droplets on surfaces requires considering the several mechanisms of osmotic regulations in the organisms. Such mechanisms are known in bulk systems. For example, bacteria respond to elevated levels of osmolality through the intracellular accumulation of osmoprotectants such as small organic molecules and ions \cite{Sleator2002}. It is a complex process primarily through uptake of K$^+$ ions, and sometimes producing molecules such as glycine betaine, proline, glutamate, trehalose, and several other structurally related zwitterionic molecules \cite{Cayley1992, Epstein1986, Wood2001}. Particularly, K$^+$ transport is the central regulatory mechanism of osmoregulation. \textit{E. coli} possesses four constitutive low-affinity K$^+$ transport systems, namely, TrkG, TrkH, Kup, and TrkF, and an inducible high-affinity system, Kdp \cite{ Epstein1986, Wood2001, Sleator2002, Cayley1992}. When bacteria are within a confined environment and experience osmotic stress, it is important to understand how different complex mechanisms of osmoregulation help their survival. Inside the evaporating microdroplets,  bacteria also experience increased concentrations of secondary metabolites. These chemical species may be unimportant in bulk solutions and at low concentrations. However, their effect will be very significant under confinement as the concentrations can shoot up by several orders due to evaporation. 

We assume that a slow increase of, for example, osmolality, activates the compensating mechanisms such as K$^+$ transport mechanisms. On the other hand, a sudden increase of osmolality may damage the bacteria before the survival mechanisms are activated. It is also reported that the osmotic stress is coupled to other stresses such as DNA supercoiling and oxidative stresses in bacteria \cite{Higgins1988, Bhriain1989, Hsieh1991, Bianchi1999, Cheung2003, Metris2014}. Also, there can be threshold concentrations of NaCl beyond which the metabolism is affected abruptly \cite{Metris2014}.  In short, evaporation of aerosols induces salt stress and associated stresses to pathogens. Even then, pathogens survive and infect their new hosts. Therefore, it should be further explored to understand mechanisms of being viable and active creating conditions present in the aerosols.  

Our method of evaporating a cluster of microdroplets can be a platform to study the response of microorganisms to the gradient of any chemical species, for example, studying bacterial susceptibility to antibiotic concentrations. 

We can achieve more control over the method by changing the microdroplet distribution inside the cluster.  Such alterations can be made, for example, physically patterning the surface to give more separation between the microdroplets. In this way, we can control the flow of oil from the cluster.

\section{Summary}

Bioaerosols are the major transporters of many contagious diseases. Extreme physiochemical conditions such as confinement and osmotic stress occur in bioaerosols due to evaporation and condensation, affecting the viability of microorganisms within them. Nevertheless, successful transmission of many pathogens occurs. Understanding the mechanism of viability requires close observation inside the aerosols, which is challenging due to the small sizes and fast dynamics of aerosols. 

In this context, we propose a method involving evaporation of a cluster of microdroplets, generated using microfluidics.  Our method can mimic evaporating bioaerosols. Importantly, the method can study the response of microorganisms to a different rate of change of the stresses. It is possible due to the gradient in the evaporation rate for droplets at different radial positions in the cluster. 

 We characterized the evaporation behavior of the cluster of microdroplets and, as a proof of concept, studied the viability of bacteria in the droplets. Our results demonstrated that not only the salt concentration but the rate of increase of the concentration (or rate of evaporation) also influence the viability of the bacteria.  
 
In short, we proposed a method that can study the viability of pathogens in evaporating aerosols and microdroplets on fomites. The method can create gradual changes in the concentration of chemical species such as antibiotics, at different rates, in droplets. As our method is based on the widely used technique of droplet-based microfluidics, researchers can further explore our idea without much delay. We hope that our proposed idea will shed light on understanding the mechanisms of the spreading of diseases through aerosols.

\section*{Conflicts of interest}
Authors declares no conflicts of interest. \medskip

\section*{Acknowledgements}
DM acknowledges Science and Engineering Research Board (India) grant CRG/2020/003117 for supporting this work. MG thanks Karthika for help in the coding for analyses. \medskip

\section*{Author contributions}
DM and RM conceived the idea, designed  the experiments and supervised. MG and AA performed the experiments and analysis. AA performed bacteria experiments and analysis. DM made the modelling and wrote the manuscript. All authors contributed to the writing. \medskip



\balance



\bibliographystyle{rsc}

\bibliography{ref1}

\providecommand*{\mcitethebibliography}{\thebibliography}
\csname @ifundefined\endcsname{endmcitethebibliography}
{\let\endmcitethebibliography\endthebibliography}{}
\begin{mcitethebibliography}{61}
\providecommand*{\natexlab}[1]{#1}
\providecommand*{\mciteSetBstSublistMode}[1]{}
\providecommand*{\mciteSetBstMaxWidthForm}[2]{}
\providecommand*{\mciteBstWouldAddEndPuncttrue}
  {\def\EndOfBibitem{\unskip.}}
\providecommand*{\mciteBstWouldAddEndPunctfalse}
  {\let\EndOfBibitem\relax}
\providecommand*{\mciteSetBstMidEndSepPunct}[3]{}
\providecommand*{\mciteSetBstSublistLabelBeginEnd}[3]{}
\providecommand*{\EndOfBibitem}{}
\mciteSetBstSublistMode{f}
\mciteSetBstMaxWidthForm{subitem}
{(\emph{\alph{mcitesubitemcount}})}
\mciteSetBstSublistLabelBeginEnd{\mcitemaxwidthsubitemform\space}
{\relax}{\relax}

\bibitem[Jones and Brosseau(2015)]{Jones2015}
R.~M. Jones and L.~M. Brosseau, \emph{J. Occup. Environ. Med.}, 2015,
  \textbf{57}, 501--508\relax
\mciteBstWouldAddEndPuncttrue
\mciteSetBstMidEndSepPunct{\mcitedefaultmidpunct}
{\mcitedefaultendpunct}{\mcitedefaultseppunct}\relax
\EndOfBibitem
\bibitem[Tellier \emph{et~al.}(2019)Tellier, Li, Cowling, and
  Tang]{Tellier2019}
R.~Tellier, Y.~Li, B.~J. Cowling and J.~W. Tang, 2019, \textbf{19}, 1--9\relax
\mciteBstWouldAddEndPuncttrue
\mciteSetBstMidEndSepPunct{\mcitedefaultmidpunct}
{\mcitedefaultendpunct}{\mcitedefaultseppunct}\relax
\EndOfBibitem
\bibitem[Dbouk and Drikakis(2020)]{Dbouk2020}
T.~Dbouk and D.~Drikakis, \emph{Phys. Fluids}, 2020, \textbf{32}, 53310\relax
\mciteBstWouldAddEndPuncttrue
\mciteSetBstMidEndSepPunct{\mcitedefaultmidpunct}
{\mcitedefaultendpunct}{\mcitedefaultseppunct}\relax
\EndOfBibitem
\bibitem[Tang(2009)]{Tang2009}
J.~W. Tang, \emph{J. R. Soc. Interface}, 2009, \textbf{6}, S737--S746\relax
\mciteBstWouldAddEndPuncttrue
\mciteSetBstMidEndSepPunct{\mcitedefaultmidpunct}
{\mcitedefaultendpunct}{\mcitedefaultseppunct}\relax
\EndOfBibitem
\bibitem[Lever \emph{et~al.}(2000)Lever, Williams, and Bennett]{Lever2000}
M.~S. Lever, A.~Williams and A.~M. Bennett, \emph{Lett. Appl. Microbiol.},
  2000, \textbf{31}, 238--241\relax
\mciteBstWouldAddEndPuncttrue
\mciteSetBstMidEndSepPunct{\mcitedefaultmidpunct}
{\mcitedefaultendpunct}{\mcitedefaultseppunct}\relax
\EndOfBibitem
\bibitem[Lighthart and Shaffer(1997)]{Lighthart1997}
B.~Lighthart and B.~T. Shaffer, \emph{Aerosol Sci. Technol.}, 1997,
  \textbf{27}, 439--446\relax
\mciteBstWouldAddEndPuncttrue
\mciteSetBstMidEndSepPunct{\mcitedefaultmidpunct}
{\mcitedefaultendpunct}{\mcitedefaultseppunct}\relax
\EndOfBibitem
\bibitem[Majee \emph{et~al.}()Majee, Chowdhury, Pinto, Chattopadhyay, Agharkar,
  Chakravortty, and Basu]{Majee}
S.~Majee, A.~R. Chowdhury, R.~Pinto, A.~Chattopadhyay, A.~N. Agharkar,
  D.~Chakravortty and S.~Basu\relax
\mciteBstWouldAddEndPuncttrue
\mciteSetBstMidEndSepPunct{\mcitedefaultmidpunct}
{\mcitedefaultendpunct}{\mcitedefaultseppunct}\relax
\EndOfBibitem
\bibitem[Haddrell and Thomas(2017)]{Haddrell2017}
A.~E. Haddrell and R.~J. Thomas, \emph{Appl. Environ. Microbiol.}, 2017,
  \textbf{83}, 809--826\relax
\mciteBstWouldAddEndPuncttrue
\mciteSetBstMidEndSepPunct{\mcitedefaultmidpunct}
{\mcitedefaultendpunct}{\mcitedefaultseppunct}\relax
\EndOfBibitem
\bibitem[Goldberg \emph{et~al.}(1958)Goldberg, Watkins, Boerke, and
  Chatigny]{Goldberg1958}
L.~J. Goldberg, H.~M. Watkins, E.~E. Boerke and M.~A. Chatigny, \emph{Am. J.
  Epidemiol.}, 1958, \textbf{68}, 85--93\relax
\mciteBstWouldAddEndPuncttrue
\mciteSetBstMidEndSepPunct{\mcitedefaultmidpunct}
{\mcitedefaultendpunct}{\mcitedefaultseppunct}\relax
\EndOfBibitem
\bibitem[Asgharian and Moss(1992)]{Asgharian1992}
B.~Asgharian and O.~R. Moss, \emph{Aerosol Sci. Technol.}, 1992, \textbf{17},
  263--277\relax
\mciteBstWouldAddEndPuncttrue
\mciteSetBstMidEndSepPunct{\mcitedefaultmidpunct}
{\mcitedefaultendpunct}{\mcitedefaultseppunct}\relax
\EndOfBibitem
\bibitem[Verreault \emph{et~al.}(2014)Verreault, Duchaine, Marcoux-Voiselle,
  Turgeon, and Roy]{Verreault2014}
D.~Verreault, C.~Duchaine, M.~Marcoux-Voiselle, N.~Turgeon and C.~J. Roy,
  \emph{Inhal. Toxicol.}, 2014, \textbf{26}, 554--558\relax
\mciteBstWouldAddEndPuncttrue
\mciteSetBstMidEndSepPunct{\mcitedefaultmidpunct}
{\mcitedefaultendpunct}{\mcitedefaultseppunct}\relax
\EndOfBibitem
\bibitem[Han \emph{et~al.}(2013)Han, Weng, and Huang]{Han2013}
Z.~Y. Han, W.~G. Weng and Q.~Y. Huang, \emph{J. R. Soc. Interface}, 2013,
  \textbf{10}, 20130560\relax
\mciteBstWouldAddEndPuncttrue
\mciteSetBstMidEndSepPunct{\mcitedefaultmidpunct}
{\mcitedefaultendpunct}{\mcitedefaultseppunct}\relax
\EndOfBibitem
\bibitem[Zhang \emph{et~al.}(2015)Zhang, Li, Xie, and Xiao]{ZhangPE2015}
H.~Zhang, D.~Li, L.~Xie and Y.~Xiao, \emph{Procedia engineering}, 2015,
  \textbf{121}, 1365--1374\relax
\mciteBstWouldAddEndPuncttrue
\mciteSetBstMidEndSepPunct{\mcitedefaultmidpunct}
{\mcitedefaultendpunct}{\mcitedefaultseppunct}\relax
\EndOfBibitem
\bibitem[Fernandez \emph{et~al.}(2019)Fernandez, Thomas, Garton, Hudson,
  Haddrell, and Reid]{Fernandez2019}
M.~O. Fernandez, R.~J. Thomas, N.~J. Garton, A.~Hudson, A.~Haddrell and J.~P.
  Reid, \emph{J. R. Soc. Interface}, 2019, \textbf{16}, 20180779\relax
\mciteBstWouldAddEndPuncttrue
\mciteSetBstMidEndSepPunct{\mcitedefaultmidpunct}
{\mcitedefaultendpunct}{\mcitedefaultseppunct}\relax
\EndOfBibitem
\bibitem[Zhen \emph{et~al.}(2014)Zhen, Han, Fennell, and Mainelis]{Zhen2014}
H.~Zhen, T.~Han, D.~E. Fennell and G.~Mainelis, \emph{J. Aerosol Sci.}, 2014,
  \textbf{70}, 67--79\relax
\mciteBstWouldAddEndPuncttrue
\mciteSetBstMidEndSepPunct{\mcitedefaultmidpunct}
{\mcitedefaultendpunct}{\mcitedefaultseppunct}\relax
\EndOfBibitem
\bibitem[Hu and Larson(2002)]{Hu2002}
H.~Hu and R.~G. Larson, \emph{J. Phys. Chem. B}, 2002, \textbf{106},
  1334--1344\relax
\mciteBstWouldAddEndPuncttrue
\mciteSetBstMidEndSepPunct{\mcitedefaultmidpunct}
{\mcitedefaultendpunct}{\mcitedefaultseppunct}\relax
\EndOfBibitem
\bibitem[Brutin and Starov(2018)]{Brutin2018}
D.~Brutin and V.~Starov, \emph{Chemical Society Reviews}, 2018, \textbf{47},
  558--585\relax
\mciteBstWouldAddEndPuncttrue
\mciteSetBstMidEndSepPunct{\mcitedefaultmidpunct}
{\mcitedefaultendpunct}{\mcitedefaultseppunct}\relax
\EndOfBibitem
\bibitem[Laghezza \emph{et~al.}(2016)Laghezza, Dietrich, Yeomans,
  Ledesma-Aguilar, Kooij, Zandvliet, and Lohse]{Laghezza2016}
G.~Laghezza, E.~Dietrich, J.~M. Yeomans, R.~Ledesma-Aguilar, E.~S. Kooij, H.~J.
  Zandvliet and D.~Lohse, \emph{Soft Matter}, 2016, \textbf{12},
  5787--5796\relax
\mciteBstWouldAddEndPuncttrue
\mciteSetBstMidEndSepPunct{\mcitedefaultmidpunct}
{\mcitedefaultendpunct}{\mcitedefaultseppunct}\relax
\EndOfBibitem
\bibitem[Hatte \emph{et~al.}(2019)Hatte, Pandey, Pandey, Chakraborty, and
  Basu]{Hatte2019}
S.~Hatte, K.~Pandey, K.~Pandey, S.~Chakraborty and S.~Basu, \emph{J. Fluid
  Mech.}, 2019, \textbf{866}, 61--81\relax
\mciteBstWouldAddEndPuncttrue
\mciteSetBstMidEndSepPunct{\mcitedefaultmidpunct}
{\mcitedefaultendpunct}{\mcitedefaultseppunct}\relax
\EndOfBibitem
\bibitem[Wen \emph{et~al.}(2019)Wen, Kim, Shi, Wang, Man, Doi, and
  Russell]{Wen2019}
Y.~Wen, P.~Y. Kim, S.~Shi, D.~Wang, X.~Man, M.~Doi and T.~P. Russell,
  \emph{Soft Matter}, 2019, \textbf{15}, 2135--2139\relax
\mciteBstWouldAddEndPuncttrue
\mciteSetBstMidEndSepPunct{\mcitedefaultmidpunct}
{\mcitedefaultendpunct}{\mcitedefaultseppunct}\relax
\EndOfBibitem
\bibitem[Netz(2020)]{Netz2020}
R.~R. Netz, \emph{The Journal of Physical Chemistry B}, 2020, \textbf{124},
  7093--7101\relax
\mciteBstWouldAddEndPuncttrue
\mciteSetBstMidEndSepPunct{\mcitedefaultmidpunct}
{\mcitedefaultendpunct}{\mcitedefaultseppunct}\relax
\EndOfBibitem
\bibitem[Dombrovsky \emph{et~al.}(2020)Dombrovsky, Fedorets, Levashov, Kryukov,
  Bormashenko, and Nosonovsky]{Dombrovsky}
L.~A. Dombrovsky, A.~A. Fedorets, V.~Y. Levashov, A.~P. Kryukov, E.~Bormashenko
  and M.~Nosonovsky, \emph{Atmosphere}, 2020, \textbf{11}, 965\relax
\mciteBstWouldAddEndPuncttrue
\mciteSetBstMidEndSepPunct{\mcitedefaultmidpunct}
{\mcitedefaultendpunct}{\mcitedefaultseppunct}\relax
\EndOfBibitem
\bibitem[Maor \emph{et~al.}(2019)Maor, Tomer, Shifra, and Nadav]{Grinberg}
G.~Maor, O.~Tomer, S.~Shifra and K.~Nadav, \emph{eLife}, 2019, \textbf{8},
  year\relax
\mciteBstWouldAddEndPuncttrue
\mciteSetBstMidEndSepPunct{\mcitedefaultmidpunct}
{\mcitedefaultendpunct}{\mcitedefaultseppunct}\relax
\EndOfBibitem
\bibitem[Seyfert \emph{et~al.}(2021)Seyfert, Rodr{\'{i}}guez-Rodr{\'{i}}guez,
  Lohse, and Marin]{Seyfert2021}
C.~Seyfert, J.~Rodr{\'{i}}guez-Rodr{\'{i}}guez, D.~Lohse and A.~Marin,
  2021\relax
\mciteBstWouldAddEndPuncttrue
\mciteSetBstMidEndSepPunct{\mcitedefaultmidpunct}
{\mcitedefaultendpunct}{\mcitedefaultseppunct}\relax
\EndOfBibitem
\bibitem[Zhu and Wang(2017)]{Zhu2017}
P.~Zhu and L.~Wang, \emph{Lab Chip}, 2017, \textbf{17}, 34--75\relax
\mciteBstWouldAddEndPuncttrue
\mciteSetBstMidEndSepPunct{\mcitedefaultmidpunct}
{\mcitedefaultendpunct}{\mcitedefaultseppunct}\relax
\EndOfBibitem
\bibitem[Shinde \emph{et~al.}(2018)Shinde, Mohan, Kumar, Dey, Maddi, Patananan,
  Tseng, Chang, Nagai, and Santra]{Shinde2018}
P.~Shinde, L.~Mohan, A.~Kumar, K.~Dey, A.~Maddi, A.~N. Patananan, F.~G. Tseng,
  H.~Y. Chang, M.~Nagai and T.~S. Santra, 2018, \textbf{19}, 3143\relax
\mciteBstWouldAddEndPuncttrue
\mciteSetBstMidEndSepPunct{\mcitedefaultmidpunct}
{\mcitedefaultendpunct}{\mcitedefaultseppunct}\relax
\EndOfBibitem
\bibitem[Gao \emph{et~al.}(2019)Gao, Jin, Zhou, and Jiang]{Gao2019}
D.~Gao, F.~Jin, M.~Zhou and Y.~Jiang, 2019, \textbf{144}, 766--781\relax
\mciteBstWouldAddEndPuncttrue
\mciteSetBstMidEndSepPunct{\mcitedefaultmidpunct}
{\mcitedefaultendpunct}{\mcitedefaultseppunct}\relax
\EndOfBibitem
\bibitem[Matu{\l}a \emph{et~al.}(2020)Matu{\l}a, Rivello, and Huck]{Matua2020}
K.~Matu{\l}a, F.~Rivello and W.~T. Huck, 2020, \textbf{4}, 1900188\relax
\mciteBstWouldAddEndPuncttrue
\mciteSetBstMidEndSepPunct{\mcitedefaultmidpunct}
{\mcitedefaultendpunct}{\mcitedefaultseppunct}\relax
\EndOfBibitem
\bibitem[Neu{\v{z}}il \emph{et~al.}(2012)Neu{\v{z}}il, Giselbrecht,
  L{\"{a}}nge, Huang, and Manz]{Neuzil2012}
P.~Neu{\v{z}}il, S.~Giselbrecht, K.~L{\"{a}}nge, T.~J. Huang and A.~Manz, 2012,
  \textbf{11}, 620--632\relax
\mciteBstWouldAddEndPuncttrue
\mciteSetBstMidEndSepPunct{\mcitedefaultmidpunct}
{\mcitedefaultendpunct}{\mcitedefaultseppunct}\relax
\EndOfBibitem
\bibitem[Shembekar \emph{et~al.}(2016)Shembekar, Chaipan, Utharala, and
  Merten]{Shembekar2016}
N.~Shembekar, C.~Chaipan, R.~Utharala and C.~A. Merten, 2016, \textbf{16},
  1314--1331\relax
\mciteBstWouldAddEndPuncttrue
\mciteSetBstMidEndSepPunct{\mcitedefaultmidpunct}
{\mcitedefaultendpunct}{\mcitedefaultseppunct}\relax
\EndOfBibitem
\bibitem[Agresti \emph{et~al.}(2010)Agresti, Antipov, Abate, Ahn, Rowat, Baret,
  Marquez, Klibanov, Griffiths, and Weitz]{Agresti2010}
J.~J. Agresti, E.~Antipov, A.~R. Abate, K.~Ahn, A.~C. Rowat, J.~C. Baret,
  M.~Marquez, A.~M. Klibanov, A.~D. Griffiths and D.~A. Weitz, \emph{Proc.
  Natl. Acad. Sci.}, 2010, \textbf{107}, 4004--4009\relax
\mciteBstWouldAddEndPuncttrue
\mciteSetBstMidEndSepPunct{\mcitedefaultmidpunct}
{\mcitedefaultendpunct}{\mcitedefaultseppunct}\relax
\EndOfBibitem
\bibitem[Kremers \emph{et~al.}(2006)Kremers, Goedhart, van Munster, and
  Gadella]{Kremers2006}
G.~J. Kremers, J.~Goedhart, E.~B. van Munster and T.~W. Gadella,
  \emph{Biochem.}, 2006, \textbf{45}, 6570--6580\relax
\mciteBstWouldAddEndPuncttrue
\mciteSetBstMidEndSepPunct{\mcitedefaultmidpunct}
{\mcitedefaultendpunct}{\mcitedefaultseppunct}\relax
\EndOfBibitem
\bibitem[Boulos \emph{et~al.}(1999)Boulos, Pr{\'{e}}vost, Barbeau, Coallier,
  and Desjardins]{Boulos1999}
L.~Boulos, M.~Pr{\'{e}}vost, B.~Barbeau, J.~Coallier and R.~Desjardins,
  \emph{J. Microbiol. Methods}, 1999, \textbf{37}, 77--86\relax
\mciteBstWouldAddEndPuncttrue
\mciteSetBstMidEndSepPunct{\mcitedefaultmidpunct}
{\mcitedefaultendpunct}{\mcitedefaultseppunct}\relax
\EndOfBibitem
\bibitem[Pontani \emph{et~al.}(2012)Pontani, Jorjadze, Viasnoff, and
  Brujic]{Pontani2012}
L.~L. Pontani, I.~Jorjadze, V.~Viasnoff and J.~Brujic, \emph{Proc. Natl. Acad.
  Sci. U. S. A.}, 2012, \textbf{109}, 9839--9844\relax
\mciteBstWouldAddEndPuncttrue
\mciteSetBstMidEndSepPunct{\mcitedefaultmidpunct}
{\mcitedefaultendpunct}{\mcitedefaultseppunct}\relax
\EndOfBibitem
\bibitem[Gauthier \emph{et~al.}(2019)Gauthier, van~der Meer, Snoeijer, and
  Lajoinie]{Gauthier2019}
A.~Gauthier, D.~van~der Meer, J.~H. Snoeijer and G.~Lajoinie, \emph{Nat.
  Commun.}, 2019, \textbf{10}, 1--5\relax
\mciteBstWouldAddEndPuncttrue
\mciteSetBstMidEndSepPunct{\mcitedefaultmidpunct}
{\mcitedefaultendpunct}{\mcitedefaultseppunct}\relax
\EndOfBibitem
\bibitem[Sun and Weisensee(2019)]{Sun2019}
J.~Sun and P.~B. Weisensee, \emph{Soft Matter}, 2019, \textbf{15},
  4808--4817\relax
\mciteBstWouldAddEndPuncttrue
\mciteSetBstMidEndSepPunct{\mcitedefaultmidpunct}
{\mcitedefaultendpunct}{\mcitedefaultseppunct}\relax
\EndOfBibitem
\bibitem[Boreyko \emph{et~al.}(2014)Boreyko, Polizos, Datskos, Sarles, and
  Collier]{Boreyko2014}
J.~B. Boreyko, G.~Polizos, P.~G. Datskos, S.~A. Sarles and C.~P. Collier,
  \emph{Proc. Natl. Acad. Sci. U. S. A.}, 2014, \textbf{111}, 7588--7593\relax
\mciteBstWouldAddEndPuncttrue
\mciteSetBstMidEndSepPunct{\mcitedefaultmidpunct}
{\mcitedefaultendpunct}{\mcitedefaultseppunct}\relax
\EndOfBibitem
\bibitem[Kajiya \emph{et~al.}(2016)Kajiya, Wooh, Lee, Char, Vollmer, and
  Butt]{Kajiya2016}
T.~Kajiya, S.~Wooh, Y.~Lee, K.~Char, D.~Vollmer and H.~J. Butt, \emph{Soft
  Matter}, 2016, \textbf{12}, 9377--9382\relax
\mciteBstWouldAddEndPuncttrue
\mciteSetBstMidEndSepPunct{\mcitedefaultmidpunct}
{\mcitedefaultendpunct}{\mcitedefaultseppunct}\relax
\EndOfBibitem
\bibitem[Karpitschka \emph{et~al.}(2016)Karpitschka, Pandey, Lubbers, Weijs,
  Botto, Das, Andreotti, and Snoeijer]{Karpitschka2016}
S.~Karpitschka, A.~Pandey, L.~A. Lubbers, J.~H. Weijs, L.~Botto, S.~Das,
  B.~Andreotti and J.~H. Snoeijer, \emph{Proc. Natl. Acad. Sci.}, 2016,
  \textbf{113}, 7403--7407\relax
\mciteBstWouldAddEndPuncttrue
\mciteSetBstMidEndSepPunct{\mcitedefaultmidpunct}
{\mcitedefaultendpunct}{\mcitedefaultseppunct}\relax
\EndOfBibitem
\bibitem[Vassileva \emph{et~al.}(2005)Vassileva, {Van Den Ende}, Mugele, and
  Mellema]{Vassileva2005}
N.~D. Vassileva, D.~{Van Den Ende}, F.~Mugele and J.~Mellema, \emph{Langmuir},
  2005, \textbf{21}, 11190--11200\relax
\mciteBstWouldAddEndPuncttrue
\mciteSetBstMidEndSepPunct{\mcitedefaultmidpunct}
{\mcitedefaultendpunct}{\mcitedefaultseppunct}\relax
\EndOfBibitem
\bibitem[Erbil(2012)]{Erbil2012}
H.~Y. Erbil, \emph{Adv. Colloid Interface Sci.}, 2012, \textbf{170},
  67--86\relax
\mciteBstWouldAddEndPuncttrue
\mciteSetBstMidEndSepPunct{\mcitedefaultmidpunct}
{\mcitedefaultendpunct}{\mcitedefaultseppunct}\relax
\EndOfBibitem
\bibitem[Semprebon \emph{et~al.}(2017)Semprebon, McHale, and
  Kusumaatmaja]{Semprebon}
C.~Semprebon, G.~McHale and H.~Kusumaatmaja, \emph{Soft matter}, 2017,
  \textbf{13}, 101--110\relax
\mciteBstWouldAddEndPuncttrue
\mciteSetBstMidEndSepPunct{\mcitedefaultmidpunct}
{\mcitedefaultendpunct}{\mcitedefaultseppunct}\relax
\EndOfBibitem
\bibitem[Carrier \emph{et~al.}(2016)Carrier, Shahidzadeh-Bonn, Zargar, Aytouna,
  Habibi, Eggers, and Bonn]{Carrier2016}
O.~Carrier, N.~Shahidzadeh-Bonn, R.~Zargar, M.~Aytouna, M.~Habibi, J.~Eggers
  and D.~Bonn, \emph{J. Fluid Mech.}, 2016, \textbf{798}, 774--786\relax
\mciteBstWouldAddEndPuncttrue
\mciteSetBstMidEndSepPunct{\mcitedefaultmidpunct}
{\mcitedefaultendpunct}{\mcitedefaultseppunct}\relax
\EndOfBibitem
\bibitem[Li \emph{et~al.}(2020)Li, Diddens, Segers, Wijshoff, Versluis, and
  Lohse]{LiPNAS2020}
Y.~Li, C.~Diddens, T.~Segers, H.~Wijshoff, M.~Versluis and D.~Lohse,
  \emph{Proceedings of the National Academy of Sciences}, 2020, \textbf{117},
  16756--16763\relax
\mciteBstWouldAddEndPuncttrue
\mciteSetBstMidEndSepPunct{\mcitedefaultmidpunct}
{\mcitedefaultendpunct}{\mcitedefaultseppunct}\relax
\EndOfBibitem
\bibitem[Smith \emph{et~al.}(2013)Smith, Dhiman, Anand, Reza-Garduno, Cohen,
  McKinley, and Varanasi]{Smith2013}
J.~D. Smith, R.~Dhiman, S.~Anand, E.~Reza-Garduno, R.~E. Cohen, G.~H. McKinley
  and K.~K. Varanasi, \emph{Soft Matter}, 2013, \textbf{9}, 1772--1780\relax
\mciteBstWouldAddEndPuncttrue
\mciteSetBstMidEndSepPunct{\mcitedefaultmidpunct}
{\mcitedefaultendpunct}{\mcitedefaultseppunct}\relax
\EndOfBibitem
\bibitem[Schellenberger \emph{et~al.}(2015)Schellenberger, Xie, Encinas, Hardy,
  Klapper, Papadopoulos, Butt, and Vollmer]{Schellenberger2015}
F.~Schellenberger, J.~Xie, N.~Encinas, A.~Hardy, M.~Klapper, P.~Papadopoulos,
  H.~J. Butt and D.~Vollmer, \emph{Soft Matter}, 2015, \textbf{11},
  7617--7626\relax
\mciteBstWouldAddEndPuncttrue
\mciteSetBstMidEndSepPunct{\mcitedefaultmidpunct}
{\mcitedefaultendpunct}{\mcitedefaultseppunct}\relax
\EndOfBibitem
\bibitem[Daniel \emph{et~al.}(2017)Daniel, Timonen, Li, Velling, and
  Aizenberg]{Daniel2017}
D.~Daniel, J.~V. Timonen, R.~Li, S.~J. Velling and J.~Aizenberg, \emph{Nat.
  Phys.}, 2017, \textbf{13}, year\relax
\mciteBstWouldAddEndPuncttrue
\mciteSetBstMidEndSepPunct{\mcitedefaultmidpunct}
{\mcitedefaultendpunct}{\mcitedefaultseppunct}\relax
\EndOfBibitem
\bibitem[Politova \emph{et~al.}(2017)Politova, Tcholakova, and
  Denkov]{Politova2017}
N.~Politova, S.~Tcholakova and N.~D. Denkov, \emph{Colloids Surfaces A
  Physicochem. Eng. Asp.}, 2017, \textbf{522}, 608--620\relax
\mciteBstWouldAddEndPuncttrue
\mciteSetBstMidEndSepPunct{\mcitedefaultmidpunct}
{\mcitedefaultendpunct}{\mcitedefaultseppunct}\relax
\EndOfBibitem
\bibitem[Marklund and Gustafsson(2001)]{Marklund2001}
O.~Marklund and L.~Gustafsson, \emph{Proc. Inst. Mech. Eng. Part J J. Eng.
  Tribol.}, 2001, \textbf{215}, 243--259\relax
\mciteBstWouldAddEndPuncttrue
\mciteSetBstMidEndSepPunct{\mcitedefaultmidpunct}
{\mcitedefaultendpunct}{\mcitedefaultseppunct}\relax
\EndOfBibitem
\bibitem[Kreder \emph{et~al.}(2018)Kreder, Daniel, Tetreault, Cao, Lemaire,
  Timonen, and Aizenberg]{Kreder2018}
M.~J. Kreder, D.~Daniel, A.~Tetreault, Z.~Cao, B.~Lemaire, J.~V. Timonen and
  J.~Aizenberg, \emph{Phys. Rev. X}, 2018, \textbf{8}, 031053\relax
\mciteBstWouldAddEndPuncttrue
\mciteSetBstMidEndSepPunct{\mcitedefaultmidpunct}
{\mcitedefaultendpunct}{\mcitedefaultseppunct}\relax
\EndOfBibitem
\bibitem[Rosenberg \emph{et~al.}(2019)Rosenberg, Azevedo, and
  Ivask]{Rosenberg2019}
M.~Rosenberg, N.~F. Azevedo and A.~Ivask, \emph{Sci. Rep.}, 2019, \textbf{9},
  1--12\relax
\mciteBstWouldAddEndPuncttrue
\mciteSetBstMidEndSepPunct{\mcitedefaultmidpunct}
{\mcitedefaultendpunct}{\mcitedefaultseppunct}\relax
\EndOfBibitem
\bibitem[Sleator and Hill(2002)]{Sleator2002}
R.~D. Sleator and C.~Hill, \emph{FEMS Microbiol. Rev.}, 2002, \textbf{26},
  49--71\relax
\mciteBstWouldAddEndPuncttrue
\mciteSetBstMidEndSepPunct{\mcitedefaultmidpunct}
{\mcitedefaultendpunct}{\mcitedefaultseppunct}\relax
\EndOfBibitem
\bibitem[Cayley \emph{et~al.}(1992)Cayley, Lewis, and Record]{Cayley1992}
S.~Cayley, B.~A. Lewis and M.~T. Record, \emph{J. Bacteriol.}, 1992,
  \textbf{174}, 1586--1595\relax
\mciteBstWouldAddEndPuncttrue
\mciteSetBstMidEndSepPunct{\mcitedefaultmidpunct}
{\mcitedefaultendpunct}{\mcitedefaultseppunct}\relax
\EndOfBibitem
\bibitem[Epstein(1986)]{Epstein1986}
W.~Epstein, \emph{FEMS Microbiol. Lett.}, 1986, \textbf{39}, 73--78\relax
\mciteBstWouldAddEndPuncttrue
\mciteSetBstMidEndSepPunct{\mcitedefaultmidpunct}
{\mcitedefaultendpunct}{\mcitedefaultseppunct}\relax
\EndOfBibitem
\bibitem[Wood \emph{et~al.}(2001)Wood, Bremer, Csonka, Kraemer, Poolman, {Van
  der Heide}, and Smith]{Wood2001}
J.~M. Wood, E.~Bremer, L.~N. Csonka, R.~Kraemer, B.~Poolman, T.~{Van der Heide}
  and L.~T. Smith, 2001, \textbf{130}, 437--460\relax
\mciteBstWouldAddEndPuncttrue
\mciteSetBstMidEndSepPunct{\mcitedefaultmidpunct}
{\mcitedefaultendpunct}{\mcitedefaultseppunct}\relax
\EndOfBibitem
\bibitem[Higgins \emph{et~al.}(1988)Higgins, Dorman, Stirling, Waddell, Booth,
  May, and Bremer]{Higgins1988}
C.~F. Higgins, C.~J. Dorman, D.~A. Stirling, L.~Waddell, I.~R. Booth, G.~May
  and E.~Bremer, \emph{Cell}, 1988, \textbf{52}, 569--584\relax
\mciteBstWouldAddEndPuncttrue
\mciteSetBstMidEndSepPunct{\mcitedefaultmidpunct}
{\mcitedefaultendpunct}{\mcitedefaultseppunct}\relax
\EndOfBibitem
\bibitem[Bhriain \emph{et~al.}(1989)Bhriain, Dorman, and Higgins]{Bhriain1989}
N.~N. Bhriain, C.~Dorman and C.~Higgins, \emph{Molecular microbiology}, 1989,
  \textbf{3}, 933--942\relax
\mciteBstWouldAddEndPuncttrue
\mciteSetBstMidEndSepPunct{\mcitedefaultmidpunct}
{\mcitedefaultendpunct}{\mcitedefaultseppunct}\relax
\EndOfBibitem
\bibitem[Hsieh \emph{et~al.}(1991)Hsieh, Rouviere-Yaniv, and Drlica]{Hsieh1991}
L.-S. Hsieh, J.~Rouviere-Yaniv and K.~Drlica, \emph{Journal of bacteriology},
  1991, \textbf{173}, 3914--3917\relax
\mciteBstWouldAddEndPuncttrue
\mciteSetBstMidEndSepPunct{\mcitedefaultmidpunct}
{\mcitedefaultendpunct}{\mcitedefaultseppunct}\relax
\EndOfBibitem
\bibitem[Bianchi and Baneyx(1999)]{Bianchi1999}
A.~A. Bianchi and F.~Baneyx, \emph{Molecular microbiology}, 1999, \textbf{34},
  1029--1038\relax
\mciteBstWouldAddEndPuncttrue
\mciteSetBstMidEndSepPunct{\mcitedefaultmidpunct}
{\mcitedefaultendpunct}{\mcitedefaultseppunct}\relax
\EndOfBibitem
\bibitem[Cheung \emph{et~al.}(2003)Cheung, Badarinarayana, Selinger, Janse, and
  Church]{Cheung2003}
K.~J. Cheung, V.~Badarinarayana, D.~W. Selinger, D.~Janse and G.~M. Church,
  \emph{Genome research}, 2003, \textbf{13}, 206--215\relax
\mciteBstWouldAddEndPuncttrue
\mciteSetBstMidEndSepPunct{\mcitedefaultmidpunct}
{\mcitedefaultendpunct}{\mcitedefaultseppunct}\relax
\EndOfBibitem
\bibitem[Metris \emph{et~al.}(2014)Metris, George, Mulholland, Carter, and
  Baranyi]{Metris2014}
A.~Metris, S.~George, F.~Mulholland, A.~Carter and J.~Baranyi, \emph{Applied
  and environmental microbiology}, 2014, \textbf{80}, 4745\relax
\mciteBstWouldAddEndPuncttrue
\mciteSetBstMidEndSepPunct{\mcitedefaultmidpunct}
{\mcitedefaultendpunct}{\mcitedefaultseppunct}\relax
\EndOfBibitem
\end{mcitethebibliography}

\end{document}